\documentclass[10pt,journal,compsoc]{IEEEtran}
\usepackage{amsmath,amsfonts}
\usepackage{algorithmic}
\usepackage{algorithm}
\usepackage{array}
\usepackage[caption=false,font=normalsize,labelfont=sf,textfont=sf]{subfig}
\usepackage{textcomp}
\usepackage{stfloats}
\usepackage{url}
\usepackage{verbatim}
\usepackage{graphicx}
\usepackage{cite}

\usepackage{mathtools}
\usepackage{optidef}
\usepackage{longtable}
\usepackage{array} 
\usepackage{booktabs}
\usepackage{tabularx} 
\usepackage[table]{xcolor}
\usepackage{colortbl}
\usepackage{paralist}
\usepackage{enumitem}

\begin{document}

\title{CODS : A Theoretical Model for Computational Design Based on Design Space}

\author{
Nan Cao,
Xiaoyu Qi,
Chuer Chen,
and~
Xiaoke Yan
\IEEEcompsocitemizethanks{
\IEEEcompsocthanksitem Nan Cao, Xiaoyu Qi, Chuer Chen, and Xiaoke Yan are with the Intelligent Big Data Visualization Lab and Shanghai Research Institute for Intelligent Autonomous Systems, Tongji University. Nan Cao is the corresponding author. Email: nan.cao@gmail.edu.cn. 
}
}

\markboth{Journal of \LaTeX\ Class Files,~Vol.~14, No.~8, August~2021}%
{Shell \MakeLowercase{\textit{et al.}}: A Sample Article Using IEEEtran.cls for IEEE Journals}

\IEEEpubid{0000--0000/00\$00.00~\copyright~2021 IEEE}

\maketitle

\begin{abstract}
We introduce CODS (\underline{C}omputational \underline{O}ptimization in \underline{D}esign \underline{S}pace), a theoretical model that frames computational design as a constrained optimization problem over a structured, multi-dimensional design space. Unlike existing methods that rely on handcrafted heuristics or domain-specific rules, CODS provides a generalizable and interpretable framework that supports diverse design tasks. Given a user requirement and a well-defined design space, CODS automatically derives soft and hard constraints using large language models through a structured prompt engineering pipeline. These constraints guide the optimization process to generate design solutions that are coherent, expressive, and aligned with user intent. We validate our approach across two domains—visualization design and knitwear generation—demonstrating superior performance in design quality, intent alignment, and user preference compared to existing LLM-based methods. CODS offers a unified foundation for scalable, controllable, and AI-powered design automation. 
\end{abstract}


\section{Introduction}
\label{sec:01_introduction}
\maketitle 


\IEEEPARstart{D}esign is inherently a complex, multidimensional process involving the composition of various elements to meet functional, aesthetic, and contextual requirements. Traditionally, designers have relied on experiential knowledge and iterative exploration to navigate the vast space of possible solutions. To formalize this exploration, the concept of the design space was introduced—a structured representation of all possible design alternatives across multiple dimensions. While widely adopted in fields such as HCI, visualization, and industrial design, existing implementations often rely on handcrafted rules or domain-specific heuristics, limiting scalability and generalizability.

In this paper, we introduce CODS (\underline{C}omputational \underline{O}ptimization in \underline{D}esign \underline{S}pace), a theoretical model that frames computational design as a constrained optimization problem over a structured design space. CODS provides a generalizable formulation that is scalable, interpretable, and adaptable to diverse design domains. Specifically, we model the design process as a multi-dimensional selection task, where each dimension represents a design factor and each solution corresponds to a point in the design space. User requirements are encoded as hard and soft constraints, which are automatically derived using large language models (LLMs) through structured prompt engineering. These constraints guide the optimization process to generate design solutions that are both functionally appropriate and semantically aligned with user intent.

We validate our approach across two design domains: visualization chart generation and knitwear design. Our experiments demonstrate that CODS not only improves design quality and consistency but also enhances interpretability and control in the generation process. This work contributes a unified theoretical foundation for computational design, enabling a principled, extensible, and AI-powered approach to intelligent design automation. Specifically, the contributions are as follows:

\begin{itemize}[leftmargin=\parindent]
    \item \textbf{A Generalizable Theoretical Framework for Computational Design:} We propose a domain-independent, formalized model that frames design as a constrained optimization problem over a structured, multi-dimensional design space. This framework captures both the combinatorial nature of design and the interplay between user intent and design principles, offering a scalable and interpretable foundation for design automation.
     \item \textbf{Automatic Constraint Generation via Large Language Models:} We introduce a novel prompt engineering pipeline that uses large language models (LLMs), such as GPT-4, to automatically extract both hard and soft constraints from natural language user requirements and structured design space representations. This approach bridges semantic understanding and computational reasoning, enabling design generation that aligns with nuanced intent without manual rule encoding.
     \item \textbf{Demonstration Across Different Design Domains with Quantitative Evaluation:} We validate the proposed model in two diverse design scenarios: (1) visualization design from natural language queries, and (2) conceptual knitwear design. Through quantitative benchmarks (e.g., VisEval for visualization) and user studies (e.g., aesthetic and conceptual alignment ratings for knitwear), the paper demonstrates that CODS yields higher-quality, more controllable, and semantically coherent design outputs than existing LLM-based baselines.
\end{itemize}

\section{Related Work}
This section reviews existing studies that are most relevant to our work, including design space and intelligent design.

\subsection{Design Space}
The concept of the design space was first introduced by Mackinlay in the 1980s to formalize experiential design knowledge~\cite{mackinlay1986automating}, and was later defined as "a space of possible designs"~\cite{mackinlay1990semantic}. Halskov and Lundqvist~\cite{halskov2021filtering} defined design space as a conceptual construct that encompasses design possibilities and supports reflection on design intentions, values, and principles. In contrast, Shaw~\cite{shaw2011role} conceptualizes it as a Cartesian space in which design decisions are the dimensions, possible alternatives are values on those dimensions, and complete designs are space points. 


Since its introduction, the concept of design space has attracted wide attention and was initially developed in depth within the domains of human-computer interaction (HCI) and visualization. For example, Gajos et al.~\cite{gajos2006exploring} explored the design space of adaptive graphical user interfaces, identifying critical design choices and benefits that determine success. Schulz et al.~\cite{6634156} proposed a general design space for visualization tasks, consolidating existing taxonomies and clarifying the conceptual understanding of operations. Beyond HCI and visualization, the design space approach has been extended to a wide range of fields, including graphic design~\cite{white2011elements}, industrial design~\cite{burnap2016estimating, chong2009human}, fashion design~\cite{liu2019parametric}, architecture~\cite{alexander1977pattern}, and even natural sciences~\cite{chen2020design,singh2020mapping} and engineering~\cite{rafiee2020multi, sepulveda2021design}.

In addition to constructing design spaces, researchers have developed tools and algorithms to operationalize these frameworks. For example, Mackinlay et al.\cite{mackinlay1986automating} proposed a graphical language design space that enables automatic chart recommendation through heuristic traversal. Moritz et al.\cite{moritz2018formalizing} formalized design knowledge as logical constraints and implemented a system for automated visualization generation. More recently, Wang et al.~\cite{wang2025jupybara} presented a multi-agent system built upon a design space for actionable data analysis and storytelling. While these works demonstrate the practical utility of design spaces, they are often tailored to specific domains and rely heavily on handcrafted rules or heuristics. To enable broader applicability, we propose a generalizable computational framework that formalizes design as an optimization process over a structured design space, enabling scalable, interpretable, and domain-independent design automation.

\subsection{Intelligent Design}
Design is the process of creating artifacts by composing design elements to serve specific purposes. Achieving high-quality results often requires designers to invest significant time and cognitive effort. To ease this burden, a wide range of techniques have been proposed to support intelligent design. 
A fundamental step is design understanding, which involves perceiving individual elements fonts~\cite{bharath2017font,zhao2018modeling,chen2014large}, visual objects~\cite{carion2020end,dou2024hierarchical,jiang2021recognizing}, and other text elements~\cite{liao2017textboxes, liu2018char, cheng2018aon}, and modeling higher-level structures like layout~\cite{manandhar2021magic,hao2023relation,shi2023reverse}, visual flow~\cite{pang2016directing,lu2020exploring}, and overall style~\cite{zhao2018characterizes,huang2021visual} that reflect principles of composition and aesthetics.

While understanding interprets existing designs, the next frontier for intelligent support lies in generation. A common early approach is template-based generation~\cite{Yang2016AutomaticGO,Cui2019TexttoVizAG,Ying2022IntelligentGA}, where designs are produced by instantiating predefined templates with content. For example, Yang et al.~\cite{Yang2016AutomaticGO} introduced topic-specific templates and proposed a computational framework for automated image cropping, typography, and color optimization within template constraints. While effective in structured scenarios, template-based methods limit creative flexibility, motivating example-based generation~\cite{Qian2020RetrieveThenAdaptEA,Shi2022SupportingEA}, which derives stylistic guidance from reference designs without rigid structural constraints. One representative work is the retrieve-then-adapt~\cite{Qian2020RetrieveThenAdaptEA} pipeline, which retrieves stylistic exemplars from online infographics and adapts them to user data. Although example-based methods offer more flexibility than templates, they still rely on existing designs, limiting novelty and adaptability.

Recent advances in generative models~\cite{Rombach2021HighResolutionIS,Podell2023SDXLIL,openai_gpt4o,Dubey2024TheL3} have opened new possibilities for flexible, creative design generation. To handle the complexity of designs, emerging works have explored modular generation~\cite{Jia2023COLEAH,Inoue2024OpenCOLETR,Chen2025POSTAAG}, where different design elements are generated in a component-wise manner. Notably, COLE~\cite{Jia2023COLEAH} used fine-tuned LLMs, large multimodal models, and diffusion models, each tailored for design-aware captioning, layout planning, and the hierarchical generation of background, objects, and typography layers for graphic design. POSTA~\cite{Chen2025POSTAAG} further leveraged background diffusion, design MLLM, and ArtText diffusion for customized artistic poster generation. In parallel, recent studies have explored end-to-end generation~\cite{Zhang2025CreatiDesignAU,Wang2025DesignDiffusionHT,Chen2025PosterCraftRH} that streamline the design process and improve visual consistency. For instance, PosterCraft~\cite{Chen2025PosterCraftRH} adopted an end-to-end framework that abandons modular pipelines and rigid layouts, enabling flexible and coherent poster composition. Despite enhancing creativity, these methods often lack explicit control over design rationale. Our approach addresses this by generating structured constraints based on a design space to guide the generation process, ensuring both visual coherence and alignment with design intent.



\section{Design Modeling}

Visual design tasks, such as visualization design, is a complex multi-dimensional and multi-stage task that involves extensive domain knowledge, including design experience and principles. To systematically represent and address these design challenges, we formulate them as an optimization problem constrained within the design space. 

\subsection{Design Space}

A design space \(\mathcal{D}\) is a multidimensional space where each dimension represents a distinct design factor that needs to be considered during the design process. Formally, it is defined as:
\[
\mathcal{D} = D_1 \times D_2 \times \dots \times D_n
\]
Each dimension consists of a collection of design elements \( e_j \in D_i \), which are the selectable values within that dimension:
\[
D_i = \{ e_{i1}, e_{i2}, \dots, e_{i m_i} \}
\]
A well-structured design space satisfies two fundamental properties: (1) its dimensions are independent, meaning no elements overlap exists among them; and (2) it comprehensively encompasses all potential design solutions, capturing the full range of possible design element combinations.

To support interpretation, the design space is typically associated with a set of meta-information. This includes descriptions of the potential users (e.g., fashion designers, interior designers, or data visualization designers), as well as the meaning of each dimension and its corresponding design elements. 



\subsection{Design Solution}

Under this definition, a design solution is a specific combination of design elements that fulfills a given requirement. It corresponds to a point in the design space, represented as: 
\[
    (e_{1j_1}, e_{2j_2}, \dots, e_{n j_n})
\]
Formally, a design solution can be expressed as a binary matrix $\mathcal{X}$:
\[
\mathcal{X} \in \{0,1\}^{n \times m}
\]
where \( n \) is the number of design dimensions, and \( m \) is the maximum number of selectable elements across all dimensions. To maintain consistency, dimensions with fewer than m elements are zero-padded. In this representation, \(\mathcal{X}_{ij} = 1\) indicates that the $j$-th element in the $i$-th dimension is included in the design solution; otherwise, the value is 0.


\subsection{Modeling the Design Process}
Given a user requirement \(\mathcal{R}\), a design solution is derived by searching the design space \(\mathcal{D}\) to select appropriate elements across multiple dimensions. This process integrates both user intent and design rules, and can be formulated as a linear programming problem:
\begin{equation}
\max \sum_{k} w_k \left( \sum_{i=1}^{n} \sum_{j=1}^{m} S_{ij}^k \cdot X_{ij} \right)
\label{eq:object}
\end{equation}
subject to
\begin{equation}
\sum_{i=1}^{n} \sum_{j=1}^{m} H_{ij}^k \cdot X_{ij} = b_k, \quad \forall k \in K_H
\label{eq:constrint}
\end{equation}
where, \(\mathcal{X} \in \{0,1\}^{n \times m}\) is the binary solution matrix, where \(X_{ij} = 1\) indicates the selection of the \(j\)-th element in the \(i\)-th dimension. Each soft rule matrix \(S^k \in \{0,1\}^{n \times m}\) encodes a preference, with \(S^k_{ij} = 1\) indicating that element \(e_{ij}\) satisfies rule \(k\). The weighted sum \(\sum_{i,j} S_{ij}^k \cdot X_{ij}\) measures how well the design solution \(\mathcal{X}\) adheres to rule \(k\), where \(w_k > 0\) promotes compliance, and \(w_k < 0\) penalizes it. In most case, \(w_k\) is a hyperparameter reflecting prior experience or domain knowledge; when such information is unavailable, \(w_k\) is typically set to either \(+1\) or \(-1\). Hard constraints are defined by the matrices \(H^k \in \{0,1\}^{n \times m}\), which represent rules that must be strictly satisfied. The parameter \(b_k\) represents the number of elements selected according to rule \(k\) regarding the design requirements.


Both hard (\(H\)) and soft (\(S\)) design rules can be derived from the user requirements \(\mathcal{R}\) and the design space \(\mathcal{D}\). In the next section, we introduce a method to automatically generate these rules using a large language model (LLM, e.g, GPT) through prompt engineering:
\begin{equation}
H = \mathrm{LLM}(\mathcal{R}, \mathcal{D}), \quad S = \mathrm{LLM}(\mathcal{R}, \mathcal{D})
\end{equation}

\begin{figure}[!t]
  \centering
  \includegraphics[width=0.9\linewidth]{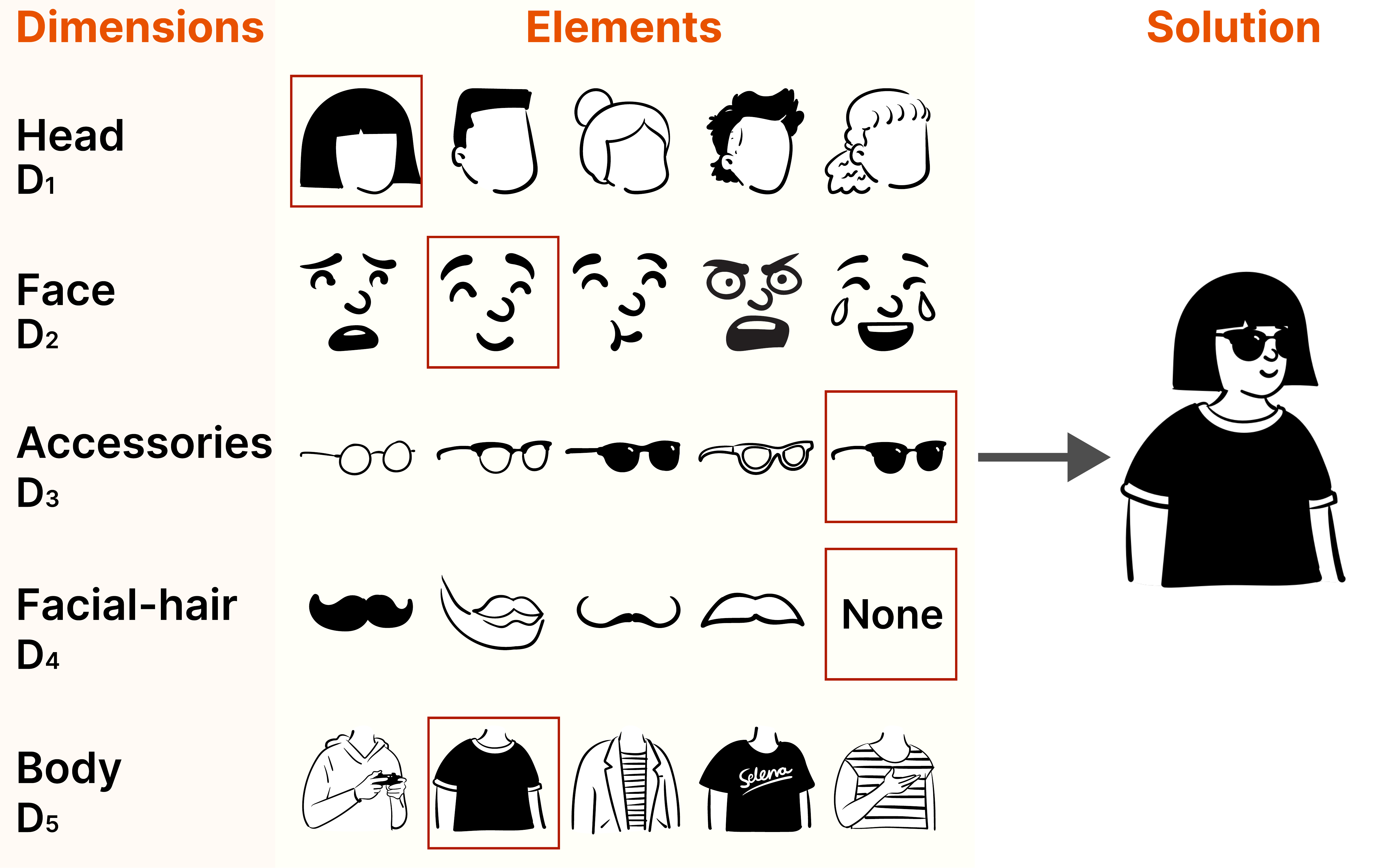}
  \caption{Design space of hand-drawn character illustrations}
  \label{fig:openpeeps}
\end{figure}

\subsection{Example: Hand-Drawn Character Illustration}
To clarify the modeling, we provide an example using hand-drawn character illustration. Fig.~\ref{fig:openpeeps} presents a design space \(\mathcal{D}\). constructed based on the Open Peeps library~\cite{openpeeps}, which contains five dimensions including head ($D_1$), face ($D_2$), accessories ($D_3$), facial-hair ($D_4$), and body ($D_5$). Each dimension consists of five representative visual elements that can be composed to generate diverse character images. 

Suppose the user requirement \(\mathcal{R}\) is \textit{"a cool and sporty girl character."} To satisfy the requirement, the design process involves both hard and soft constraints. For instance, one hard constraint restricts the \textbf{head} dimension to select a female head (e.g., \texttt{woman bangs black}), and another requires the \textbf{facial-hair} dimension to select \texttt{none}. Each hard constraint corresponds to a separate matrix:

\begin{equation}
H_1 = 
\begin{bmatrix}
1 & 0 & 1 & 0 & 1 \\
0 & 0 & 0 & 0 & 0 \\
0 & 0 & 0 & 0 & 0 \\
0 & 0 & 0 & 0 & 0 \\
0 & 0 & 0 & 0 & 0
\end{bmatrix}, \quad
H_2 = 
\begin{bmatrix}
0 & 0 & 0 & 0 & 0 \\
0 & 0 & 0 & 0 & 0 \\
0 & 0 & 0 & 0 & 0 \\
0 & 0 & 0 & 0 & 1 \\
0 & 0 & 0 & 0 & 0
\end{bmatrix}
\end{equation}

Soft constraints reflect preferences inferred from the requirement. For example, the \textbf{face} dimension prefers \texttt{calm} to convey a cool attitude. Another constraint encourages selecting \texttt{sunglasses} in \textbf{accessories} and \texttt{sporty tee} in \textbf{body} to enhance the sporty appearance. Each soft constraint is represented as a separate matrix:

\begin{equation}
S_1 = 
\begin{bmatrix}
0 & 0 & 0 & 0 & 0 \\
0 & 1 & 0 & 0 & 0 \\
0 & 0 & 0 & 0 & 0 \\
0 & 0 & 0 & 0 & 0 \\
0 & 0 & 0 & 0 & 0
\end{bmatrix}, \quad
S_2 = 
\begin{bmatrix}
0 & 0 & 0 & 0 & 0 \\
0 & 0 & 0 & 0 & 0 \\
0 & 0 & 0 & 0 & 1 \\
0 & 0 & 0 & 0 & 0 \\
0 & 1 & 0 & 0 & 0
\end{bmatrix}
\end{equation}

Solving the objective function (Eq.\ref{eq:object}) subject to the hard constraints (Eq.\ref{eq:constrint}) yields a design solution matrix:
\begin{equation}
\mathcal{X} =
\begin{bmatrix}
1 & 0 & 0 & 0 & 0 \\
0 & 1 & 0 & 0 & 0 \\
0 & 0 & 0 & 0 & 1 \\
0 & 0 & 0 & 0 & 1 \\
0 & 1 & 0 & 0 & 0
\end{bmatrix}
\end{equation}

In tuple form, the design solution can be expressed as::
\begin{equation}
    (e_{11},e_{22},e_{35},e_{45},e_{52})
\end{equation}
This solution corresponds to a character with a bob haircut, calm face, sunglasses, no facial hair, and a sporty tee, effectively capturing the intended look.

\section{Constraint Generation}
\label{sec:constraint_generation}
In this section, we present the methods used to generate constraints from the input design space and user requirements by leveraging LLM.

\begin{figure}[th]
  \centering
  \includegraphics[width=\linewidth]{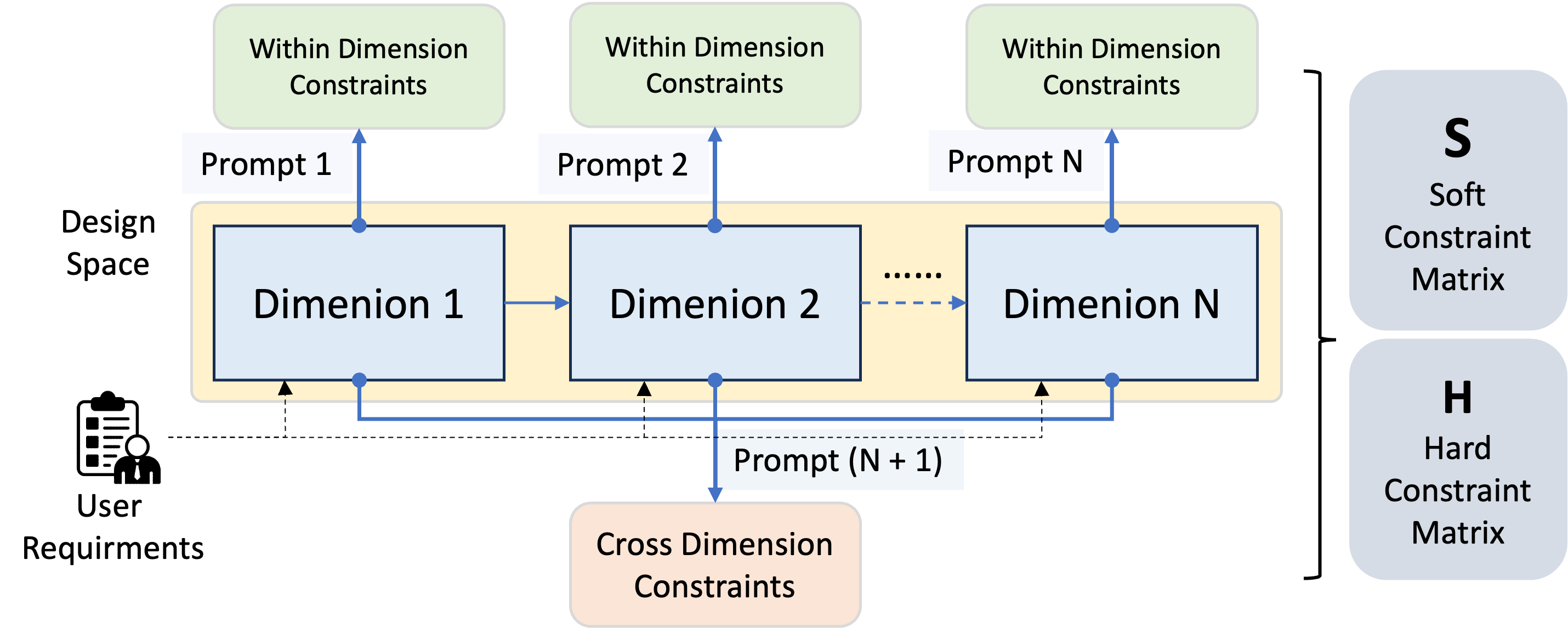}
  \caption{Constraint Generation Pipeline.}
  \label{fig:pipeline}
\end{figure}

\begin{figure*}[!ht]
  \centering
  \includegraphics[width=\linewidth]{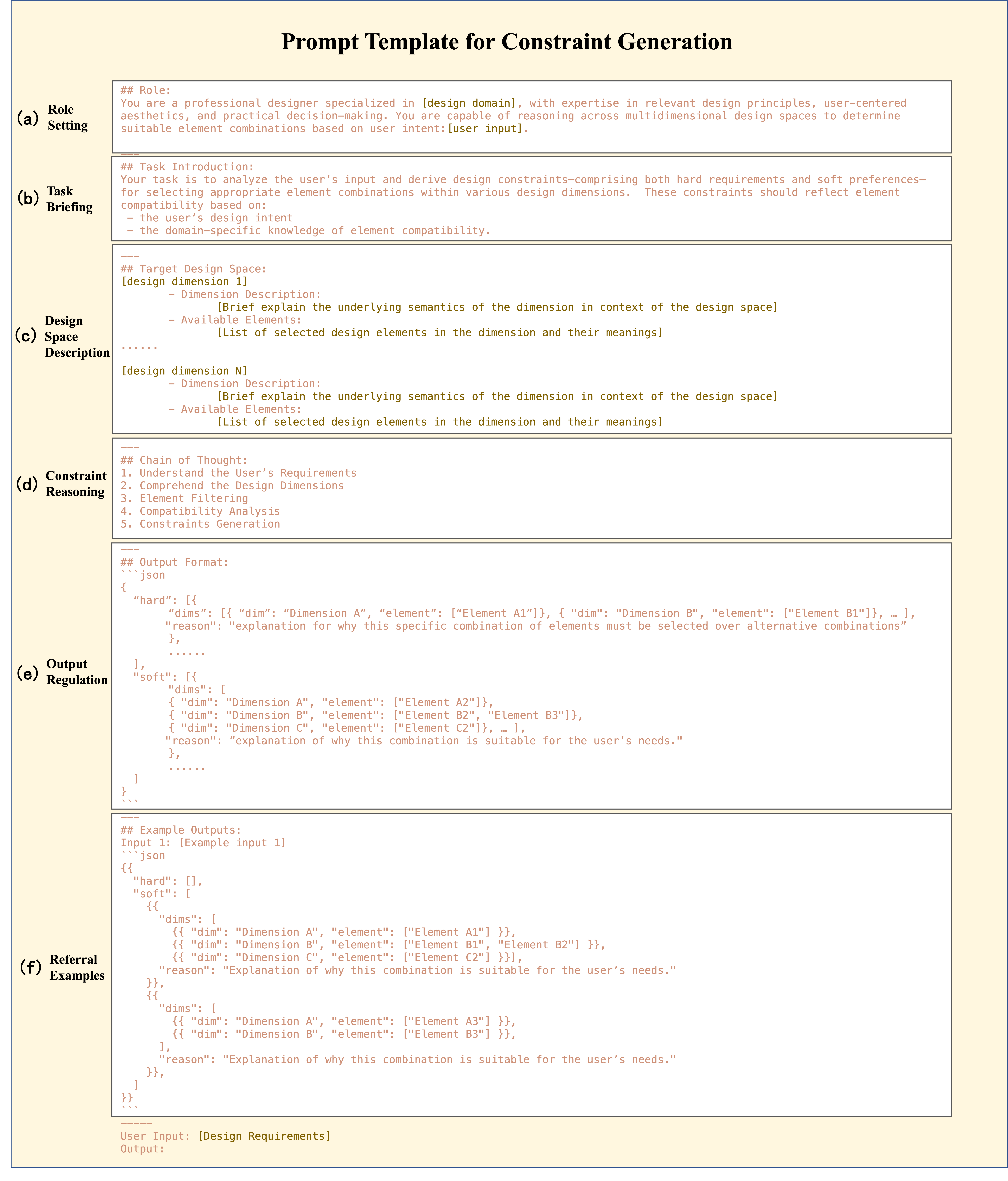}
  \vspace{-20pt}
  \caption{The prompt template for constraint generation consists of six components: (a) role setting, (b) task briefing, (c) design space description, (d) constraint reasoning, (e) output regulation, and (f) referral examples}
  \vspace{-5pt}
  \label{fig:prompt}
\end{figure*}

\subsection{Overview}
As illustrated in Fig.~\ref{fig:pipeline}, a series of prompts is constructed and executed sequentially (from prompt 1 to prompt N+1) using GPT-4 to support constraint generation. Specifically, intra-dimensional constraints are first generated by preparing a prompt for each dimension individually. Subsequently, cross-dimensional constraints are generated using a prompt that includes all relevant dimensions and their associated elements aligned with the user's requirements. During this process, prompts are carefully designed to guide the LLM's reasoning, encouraging it to estimate and establish the relationships between elements either within a single dimension or across multiple dimensions. The resulting constraints define the compatibility or incompatibility among design elements regarding users' requirements.


Both soft and hard constraints are generated simultaneously in the above process. Soft constraints typically express design preferences or desirable characteristics that can be flexibly adjusted, while hard constraints define fundamental design rules intrinsic to the design space \(\mathcal{D}\) or directly inferred from the user requirements \(\mathcal{R}\), and must not be violated. For example, if the user specifies ``the illustration of a pretty girl,'' the system must ensure that hairstyle and clothing elements do not exhibit masculine characteristics.

Finally, all the constraints, both generated or selected, are parsed into two separate matrices: \(\mathcal{S}\) for soft constraints and \(\mathcal{H}\) for hard constraints, as defined in equation~(\ref{eq:object}) and equation~(\ref{eq:constrint}). These matrices serve as the input for the subsequent optimization process.




\subsection{Prompt Engineering}
Given the user requirements and the associated design space \(\mathcal{D}\), we construct a series of prompts to extract constraints within each design dimension and across different dimensions based on a template that comprises six components: (a) role setting, (b) task briefing, (c) design space description, (d) constraint reasoning, (e) output regulation, and (f) referral examples (Fig.~\ref{fig:prompt}):

\textbf{Role Setting (Fig.~\ref{fig:prompt}(a)).} Our prompt begins by defining the LLM’s role as a designer within a specific domain (e.g., fashion, interior, or data visualization), derived from the meta-information of the design space. This role-setting strategy has been shown to be effective in various prompt engineering tasks, as it helps contextualize the LLM's reasoning within the relevant design domain.

\textbf{Task Briefing (Fig.~\ref{fig:prompt}(b)).} 
Next, the task is briefly described: to analyze the user's input and derive appropriate design constraints, including both hard requirements and soft preferences, for selecting suitable design elements within each individual dimension and across multiple dimensions of the design space. These constraints are logically grounded in the user's requirements and serve to guide the selection process by narrowing the available options to those that best align with the intended design goals.

\textbf{Design Space Description (Fig.~\ref{fig:prompt}(c)).} In this part, detailed descriptions of the design dimensions in the given design space are provided, including definitions and the corresponding set of available design elements. This information enhances the model’s understanding of the design space and ensures more informed and context-aware reasoning. 

\textbf{Constraint Reasoning (Fig.~\ref{fig:prompt}(d)).}  
With the preceding contextual information, we structure the constraint generation process into a series of systematic reasoning steps that form a chain of thought to steer the reasoning of the large language model (LLM). By combining a series of carefully designed short prompts through the pipeline, this approach enhances reasoning capabilities through intermediate steps, making the design decision process more transparent and interpretable: 

\textit{\textbf{Step 1: Understand the User’s Requirements.}} Analyze the user's input to extract expectations, needs, and design intentions. Consider design styles, emotional tone, cultural references, target demographics, functional objectives, and stylistic cues.

\textit{\textbf{Step 2: Comprehend the Design Space.}} Clarify the role of the given design dimensions within the overall design space. Identify the relationships between different dimensions. Interpret the semantics and practical implications of the available elements in the dimension. 

\textit{\textbf{Step 3: Element Filtering.}} Reason about the distinguishing factors among design elements within each dimension. Identify the key attributes that differentiate one element from another and infer the underlying principles that guide element selection. Apply domain expertise to interpret how user needs can be effectively mapped to these attributes, establishing conceptual links between user intent and available design options. Based on this reasoning, filter and retain the elements that best satisfy the design requirements and most closely align with the user's intent.

\textit{\textbf{Step 4: Compatibility Analysis.}} Assessing the compatibility of the filtered design elements within/across different dimensions. For each possible pair or group of elements, evaluate their relationship based on the following criteria: semantic alignment, visual harmony, stylistic coherence, emotional resonance, functional complementarity, and their relevance to the design requirements given by users. Use relevant domain knowledge to support the reasoning and identify: (1) element combinations that are suitable and recommended for selection together; (2) element combinations that are incompatible and should not be selected together. Provide brief reasoning for each decision to explain why the elements are compatible or incompatible.

\textit{\textbf{Step 5: Constraints Generation.}} Derive both the hard and soft constraints from the selected elements within or across different dimensions. Specifically, within a single dimension, \textit{hard constraints} are defined as design elements that are explicitly specified by the user's input, reflecting mandatory requirements. In contrast, \textit{soft constraints} represent alternative elements that are consistent with the user's intent but are not strictly required for the design solution. Across multiple dimensions, \textit{hard constraints} identify element combinations that must either be selected together or explicitly avoided, based on logical or functional dependencies. Conversely, \textit{soft constraints} describe preferred but non-mandatory associations between elements, indicating desirable combinations or separations that align with the user's goals while allowing flexibility during solution generation.

\textbf{Output Regulation (Fig.~\ref{fig:prompt}(e)).} This part of the prompt provides the desired JSON format of the generated constraints.

\textbf{Referral Examples (Fig.~\ref{fig:prompt}(f)).}
A small set of input-output pairs is also provided in the template as a reference for few-shot learning. This approach enables the model to internalize the underlying patterns and reasoning strategies for constraint generation from a limited number of high-quality examples, enhancing its accuracy and consistency when generating constraints for new inputs.

It is worth noting that although the same template and general reasoning steps are followed, generating constraints within and across dimensions relies on different reasoning instructions. Intra-dimensional constraint generation focuses on selecting elements within a single dimension that satisfy the user's requirements, whereas cross-dimensional constraint generation emphasizes establishing relationships between elements across multiple dimensions. 

Next, we demonstrate the effectiveness of the proposed technique through two real-world design applications: visualization chart generation based on users' data queries and knitwear design based on users' requirements.

\section{Visualization Design}
In this section, we validate the proposed method in a visualization design scenario. We first define a visualization design space, followed by a description of the implementation details. We then showcase the generated results and quantitatively evaluate the method's performance.

\begin{table*}[th]
\centering
\caption{The design space of visualization}
\label{tab:chart_design_space}
\begin{tabular}{m{0.12\textwidth} m{0.56\textwidth} m{0.24\textwidth}}
\toprule
\textbf{Dimension} & \textbf{Description} & \textbf{Design Elements} \\
\midrule
Mark Type & Specifies the fundamental graphical primitives that constitute the visual elements of the chart and determine the chart type.  & \textit{bar}, \textit{line}, \textit{point}, \textit{pie} \\
\midrule
X & Specifies the mapping of a data field to the spatial channel defined by the x-axis.& \textit{[Data Fields]} \\
\midrule
Y & Specifies the mapping of a data field to the spatial channel defined by the y-axis.& \textit{[Available Data Fields]}, \textit{none} \\
\midrule
Color & Specifies the mapping of a data field to the color channel. & \textit{[Data Fields]}, \textit{none} \\
\midrule
Size & Specifies the mapping of a data field to the size channel. & \textit{[Data Fields]}, \textit{none} \\
\midrule
Group By & Specifies the categorical data field by which the data records are grouped. & \textit{[Categorical Data Fields]}, \textit{none} \\
\midrule
Aggr. Method on X & Specifies the method for aggregating the field mapping to X when data records are grouped. & \textit{average}, \textit{sum}, \textit{count}, \textit{min}, \textit{max}, \textit{none} \\
\midrule
Aggr. Method on Y & Specifies the method for aggregating the field mapping to Y when data records are grouped. & \textit{average}, \textit{sum}, \textit{count}, \textit{min}, \textit{max}, \textit{none} \\
\midrule
Aggr. Method on Size & Specifies the method for aggregating the field mapping to Size when data records are grouped. & \textit{average}, \textit{sum}, \textit{count}, \textit{min}, \textit{max}, \textit{none} \\
\midrule
Sort & Specifies the data field by which the records are ordered. & \textit{[Data Fields]}, \textit{none} \\
\midrule
Order & Specifies the sorting direction. & \textit{ascending}, \textit{descending}, \textit{none} \\
\bottomrule
\end{tabular}

\vspace{0.5em}
\raggedright
\textit{[Data Fields]} indicates all available data fields in the input dataset.
\end{table*}

\subsection{Design Space}

Drawing on well established knowledge in information visualization~\cite{satyanarayan2016vega,6634156,mackinlay1986automating}, we construct a visualization design space comprising eleven dimensions (Table~\ref{tab:chart_design_space}), with certain dimension elements dynamically generated from the user-uploaded dataset. The dimensions are detailed as follows:


\textbf{Mark Type} specifies the fundamental graphical primitives that constitute the visual elements of a chart. It determines the geometric representation of data items and thereby defines the chart type. Our framework supports four mark types: \textit{bar}, \textit{line}, \textit{point}, and \textit{pie}.


\textbf{X / Y} specify the mapping of data fields to spatial channels on the x- and y-axis. The elements correspond to fields in the uploaded dataset. Notably, \textit{Y} can be set to \textit{none}, indicating that the y-axis represents aggregated values rather than a raw data field.

\textbf{Color / Size} specify the mapping of data fields to color and size channels, enabling the visual encoding of additional data attributes. Elements for each dimension include data fields or \textit{none} to disable the channel.

\textbf{Group By} specifies the data field used to group records. Available elements include the categorical fields or \textit{none}, where \textit{none} indicates that no grouping is performed.

\textbf{Aggregation Method on X/Y/Size} specify the aggregation methods applied to the data fields mapped to the X, Y, or size channels when the above grouping is enabled. Supported methods include \textit{average}, \textit{sum}, \textit{count}, \textit{min}, \textit{max}, and \textit{none} (no aggregation).

\textbf{Sort} specifies the data field by which the records are ordered. Available elements include data fields or \textit{none}.

\textbf{Order} specifies the sorting direction: \textit{ascending}, \textit{descending}, or \textit{none} (i.e., when no sorting field is defined).



\subsection{Implementation}
Given a user-uploaded dataset and a data query representing the user requirement, we first dynamically construct a visualization design space based on the dataset. Using the design space and the user requirement, we prompt GPT-4~\cite{achiam2023gpt} to generate design constraints. These constraints are then formulated into the linear programming problem defined in Section 3.3 and solved using the SCIP solver~\cite{achterberg2009scip} to derive a design solution. Finally, the resulting solution is rendered into a visualization using the Seaborn library~\cite{waskom2021seaborn}.





\subsection{Evaluation}
We evaluate our approach by presenting generated visualizations and conducting a performance comparison with other LLM-based visualization methods.

\subsubsection{Generation Results}

\begin{figure*}[th]
  \centering
  \includegraphics[width=\linewidth]{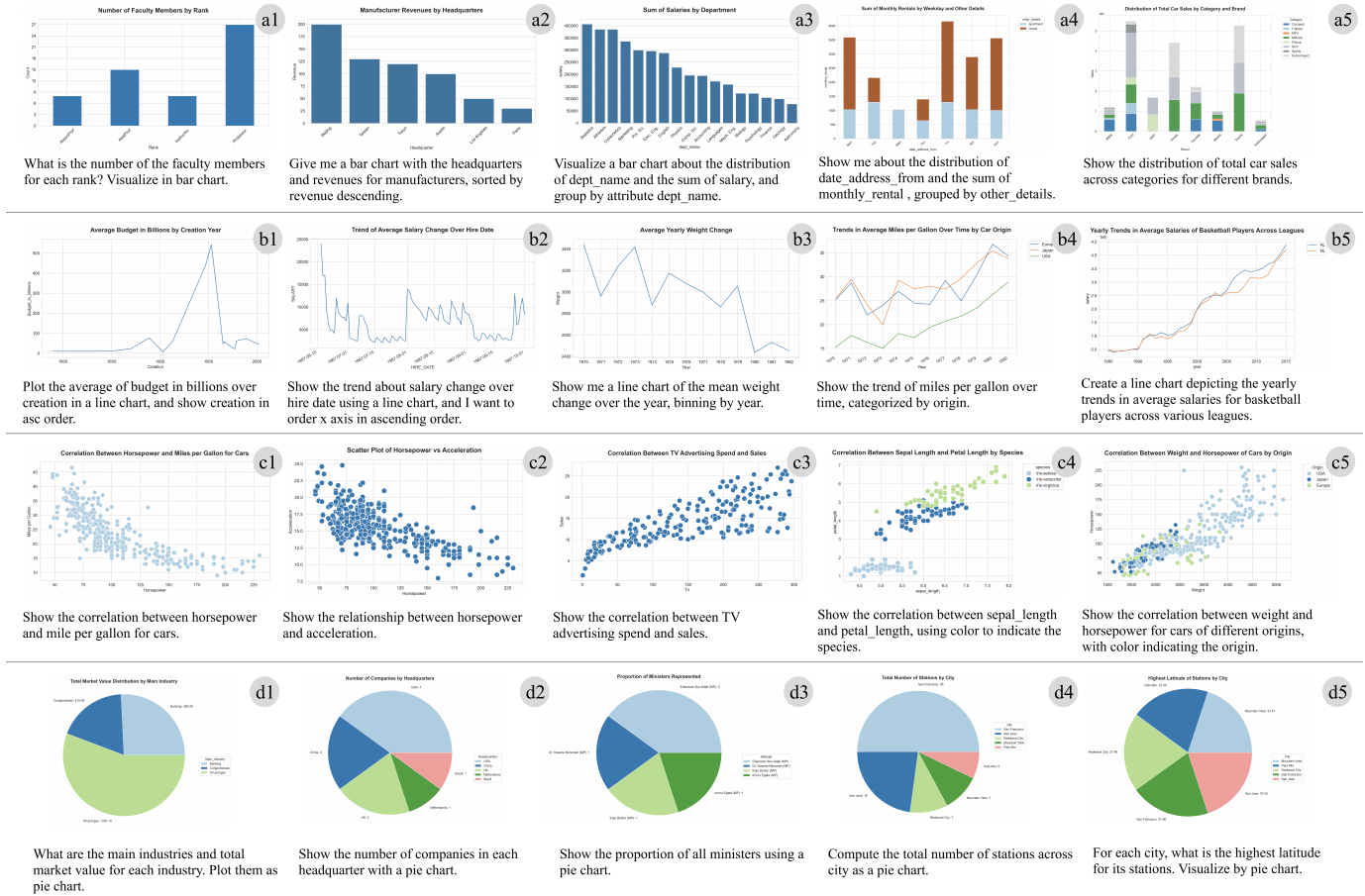}
  \caption{Generated visualization results with (a1)–(a5) displaying \textit{bar} charts, (b1)–(b5) \textit{line} charts, (c1)–(c5) \textit{point} charts, (d1)–(d5) \textit{pie} charts.}
  \label{fig:visualization_result}
\end{figure*}


We present a variety of visualizations generated by our approach, as shown in Fig
.~\ref{fig:visualization_result}. For example, Fig.~\ref{fig:visualization_result}-(a4) shows the stacked bar chart generated for a query \textit{"Show me about the distribution of 'date address from' and the sum of 'monthly rental', grouped by other details."}. Our approach selects \texttt{bar} in the \textbf{mark type} dimension, maps relevant fields to the \textbf{x}, \textbf{y}, and \textbf{color} dimensions, and applies \texttt{sum} in \textbf{aggregation method on the y} dimension, successfully capturing the intended data distribution. Fig.~\ref{fig:visualization_result}-(c1) shows a scatterplot generated for the query \textit{"Show the correlation between weight and mile per gallon for cars."}. Our method selects \texttt{point} as the \textbf{mark type} to represent individual data entries and maps the fields \texttt{weight} and \texttt{mpg} to the \textbf{x} and \textbf{y} axes, effectively revealing the correlation pattern described in the query. These examples demonstrate that our approach can semantically interpret user requirement, make informed design decisions within the visualization design space, and generate corresponding visualizations.

\subsubsection{Performance Evaluation}

\begin{table*}[t]
\centering
\caption{Performance of different methods on VisEval benchmark. \textbf{Bold} values indicate the best-performing results among GPT-3.5-based methods for each metric. \textcolor{red}{\textbf{Red bold}} values indicate the best results across all methods. The last row reports the performance of our approach on a filtered subset that excludes queries requiring data filtering.}
\label{tab:seaborn-results}
\begin{tabular}{lccccc}
\toprule
\textbf{Method} & \textbf{Invalid Rate }$\downarrow$ & \textbf{Illegal Rate }$\downarrow$ & \textbf{Pass Rate }$\uparrow$ & \textbf{Readability Score }$\uparrow$ & \textbf{Quality Score }$\uparrow$ \\
\midrule
CoML4VIS(GPT-3.5)  & 6.95\%  & 32.02\%  & 61.03 \%  & 3.66 & 2.27 \\
LIDA(GPT-3.5)  & 12.69\% & 35.54\%  & 51.77\%  & 3.39 & 1.77 \\
Chat2VIS(GPT-3.5)   & 3.65\%  & \textcolor{red}{\textbf{31.05\%}} & \textbf{65.30\%}  & 3.04 & 2.02 \\
Our Approach(GPT-3.5) &  \textbf{0.77}\%  & 34.06\%  & 65.17\%  & \textbf{3.73} & \textbf{2.46} \\
Our Approach(GPT-4) &  \textcolor{red}{\textbf{0.67\%}}  & 31.14\%  & \textcolor{red}{\textbf{67.67\%}}  & \textcolor{red}{\textbf{3.80}} & \textcolor{red}{\textbf{2.61}} \\
\midrule
Our Approach(GPT-4, filter excluded) &  0.0\%  & 26.39\%  & 73.61\%  & 3.98 & 2.98 \\
\bottomrule
\end{tabular}
\end{table*}

We evaluate the visualizations produced by our approach using the VisEval~\cite{chen2024viseval} framework and compare them with other LLM-based visualization approaches.



\textbf{Baselines}. We compare our approach with three visualization generation methods powered by LLMs: Chat2VIS~\cite{maddigan2023chat2vis}, LIDA\cite{dibia2023lida}, and CoML4VIS~\cite{chen2024viseval}. Chat2VIS uses LLMs with tailored prompts to convert natural language into Python visualization code. LIDA employs a multi-stage pipeline where an LLM interprets data semantics, enumerates visualization goals, and generates charts. CoML4VIS adapts CoML~\cite{zhang2023mlcopilot}, an LLM-based coding assistant, by refining prompts to focus on visualization generation. We adopt evaluation results reported in the VisEval benchmark~\cite{chen2024viseval}, where all baselines were evaluated using GPT-3.5 as the underlying model. For consistency, we also use GPT-3.5 in our approach for comparison, and further evaluate our approach with GPT-4 to assess the potential improvements enabled by a stronger language model.



\textbf{Dataset and Metrics}. We use the VisEval~\cite{chen2024viseval} benchmark, which includes a dataset of 2,524 natural language queries covering 146 databases, each paired with a ground-truth visualization. For our evaluation, we use the subset with single-table data, consisting of 689 queries. All methods are evaluated using the metrics provided by VisEval:

\begin{enumerate}
\item \textbf{Invalid Rate}: The percentage of code that crashes or fails to execute in a sandbox environment;
\item \textbf{Illegal Rate}: The proportion of visualizations that do not match the query;
\item \textbf{Pass Rate}: The rate of code that executes successfully and generates a visualization consistent with the query;
\item \textbf{Readability Score}: A measure of the visualization’s clarity, including text overflow and canvas layout;
\item \textbf{Quality Score}: A holistic assessment of the generated visualization.
\end{enumerate}


\textbf{Results}. The quantitative evaluation results are summarized in Table~\ref{tab:seaborn-results}. Among all methods evaluated with the GPT-3.5 model, our approach achieves the lowest invalid rate (0.77\%), primarily due to constrained optimization, which effectively reduces code execution failures caused by visualization rule violations. Furthermore, our approach achieves the highest readability scores (3.73) and quality scores (2.46), reflecting its strength in enhancing visual clarity and overall design quality. In terms of pass rate, our method (65.17\%) performs comparably to the best baseline (65.30\%), indicating its capability to generate valid and legal visualizations. 

However, our approach shows moderate performance on illegal rate, which may be partly attributed to the design space’s focus on visual design rather than data analysis, limiting its ability to handle complex query requirements involving data filtering. To investigate the issue, we conducted an additional evaluation on a filtered subset (\textit{n} = 478) that excludes queries requiring data filtering. Using the same GPT-4 model, the illegal rate on this subset decreased from 31.14\% to 26.39\%, suggesting that complex data operations are a primary source of illegal outputs. This filtered evaluation also yields the best performance across all metrics, demonstrating the method’s potential when complex data operations are excluded.

In addition, when using the GPT-4 model, our approach shows further improvements across all metrics, obtaining the best performance in invalid rate, pass rate, readability score and quality score, suggesting that enhanced language understanding can further strengthen the effectiveness of our method.

\section{Knitwear Design}
In this section, we apply the proposed method to the domain of knitwear design, demonstrating its applicability across diverse design scenarios. We begin by constructing a design space for knitwear that reflects the unique features of textile patterns and structures. Next, we describe the implementation details of our approach in this context. Finally, we present the generated knitwear designs and conduct a user study comparing our approach with a baseline method.

\subsection{Design Space}
Building upon prior studies within this domain~\cite{motta2019designing,petre2006complexity,jones2021computational}, we define a knitwear design space comprising six design dimensions. Each dimension includes a set of design elements, with counts shown in parentheses: Garment Types (28), Surface Patterns (20), Knitting Techniques (28), Aesthetic Styles (30), Color Palettes (40), and Visual Motifs (36). Table~\ref{tab:design_space_knitwear} lists representative elements, with explanations of the dimensions provided below.

\begin{table*}[t]
\centering
\caption{The design space for knitwear}
\label{tab:design_space_knitwear}
\begin{tabular}{m{0.15\textwidth} m{0.46\textwidth} m{0.32\textwidth}}
\toprule
\textbf{Dimension} & \textbf{Description} & \textbf{Design Elements} \\
\midrule
Garment Type & Defines the structural form and silhouette of the knitwear. & \textit{Hoodie}, \textit{Jacket}, \textit{Turtleneck sweater}, \textit{Henley shirt}, \textit{A-line dress}, \textit{Off-shoulder dress}, etc. \\
\midrule
Surface Pattern & Describes the surface textures and decorative patterns of knitted fabrics. & \textit{Ribbed knit detail}, \textit{Jacquard weave pattern}, \textit{Tweed knit surface}, \textit{Trellis knit texture}, etc. \\
\midrule
Knitting Technique & Refers to the knitting techniques used in fabric construction. Given the inherent softness and elasticity of knitted fabrics, variations in technique yield diverse structural characteristics and material aesthetics. & \textit{Herringbone stitch}, \textit{Chevron stitch}, \textit{Seed stitch}, \textit{Moss stitch}, \textit{Lace stitch}, \textit{Rib stitch}, \textit{Brioche stitch}, \textit{Waffle stitch} etc. \\
\midrule
Aesthetic Style & Captures the overall aesthetic identity of the knitwear design, reflecting culturally distinctive and visually recognizable stylistic paradigms. & \textit{Nordic Folk}, \textit{Vintage-inspired}, \textit{Bohemian Crochet}, \textit{Minimalist}, etc. \\
\midrule
Color Palette & Defines the color composition and thematic palettes in knitwear design. & \textit{Pastel color}, \textit{Tropical color}, \textit{Vintage floral color}, \textit{Architectural gray color}, etc. \\
\midrule
Visual Motif & Represents the conceptual theme and abstract inspiration underlying the design, translating intangible ideas into tangible visual forms to guide the garment’s aesthetic direction and thematic expression. & \textit{Cloud silhouette}, \textit{Marshmallow texture}, \textit{Magic forest}, \textit{Oil paint touch}, \textit{Geometric blocks}, \textit{Pixel shapes}, \textit{Mechanical gears}, etc. \\
\bottomrule
\end{tabular}
\end{table*}

\textbf{Garment Type} defines the structural form and silhouette of the knitwear. 

\textbf{Surface Pattern} describes the surface textures and decorative patterns of knitted fabrics, which significantly contribute to the garment's visual aesthetics. 

\textbf{Knitting Technique} indicates the methods employed in fabric construction. Given the inherent softness and elasticity of knitted fabrics, variations in technique yield diverse structural characteristics and material aesthetics. 

\textbf{Aesthetic Style} defines the overall aesthetic identity of the knitwear design, reflecting culturally distinctive and visually recognizable stylistic paradigms. 

\textbf{Color Palette} describes the color composition and emotional tone used in knitwear design. 

\textbf{Visual Motif} represents the conceptual theme and abstract inspiration behind the design. It translates intangible concepts into tangible visual forms, guiding the garment’s aesthetic direction and thematic expression. 


\subsection{Implementation}
We leverage GPT-4 to generate constraints over the knitwear design space based on user requirements. These constraints define a linear programming problem, which we solve using the SCIP~\cite{achterberg2009scip} solver to obtain a solution. The selected design elements from the solution are extracted as keywords and composed into a textual prompt, which is input into the advanced text-to-image model Midjourney~\cite{midjourney} to generate knitwear renderings.

\subsection{Evaluation}
In this section, we evaluate the effectiveness of our proposed method by comparing the quality of knitwear renderings generated by our approach and a baseline method. The baseline uses GPT-4 to directly convert user requirements into text-to-image prompts, which are fed into Midjourney to generate knitwear renderings. We first present eight groups of generation results from both methods for visual comparison. Then, we conduct a questionnaire-based user study to assess their quality.

\subsubsection{Generation Results}

\begin{figure*}[th]
  \centering
  \includegraphics[width=\linewidth]{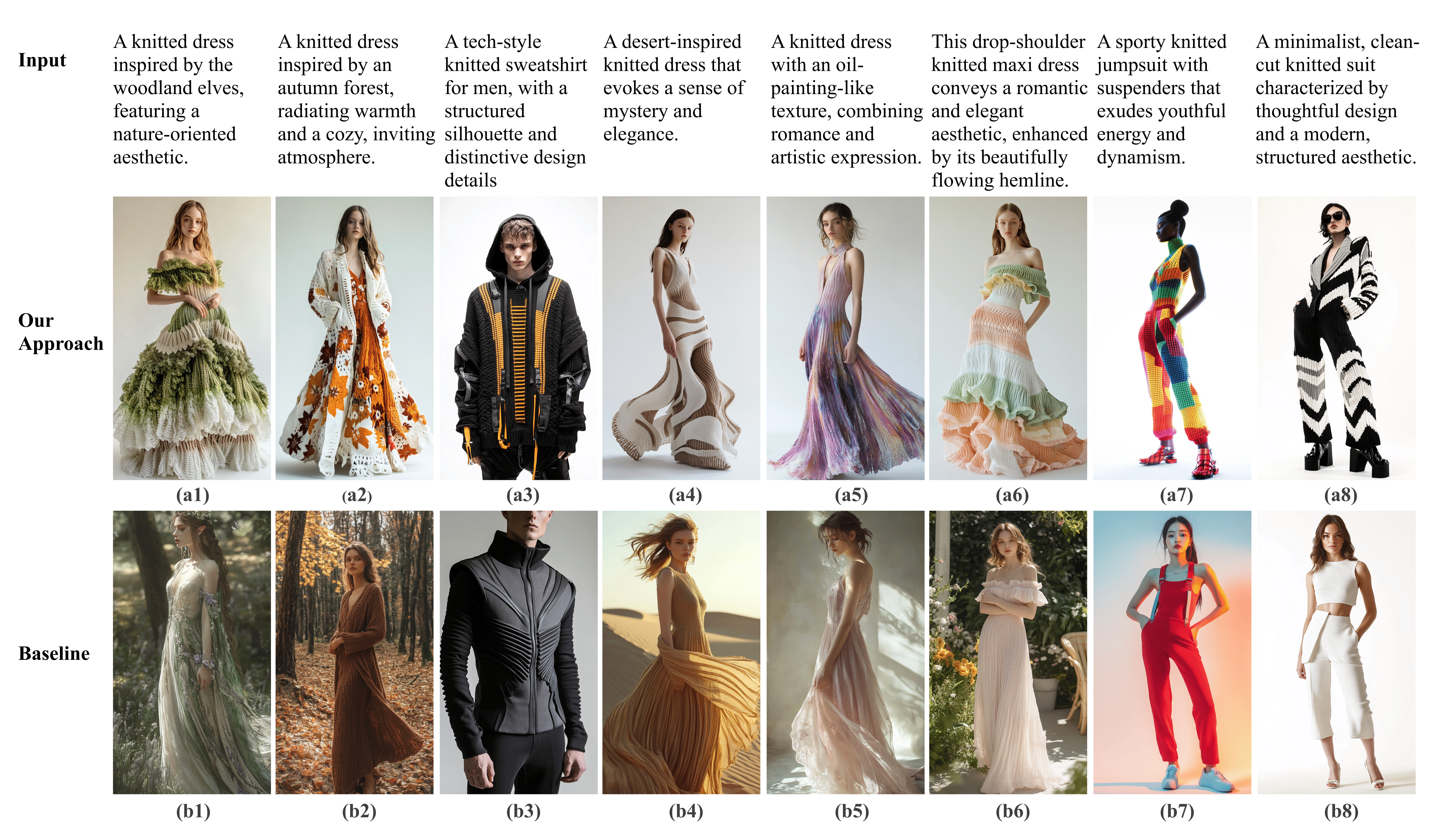}
  \caption{The results of generated knitwear renderings, with (a1-a8) generated by our approach and (b1-b8) generated by baseline approach.} 
  \label{fig:knitwear_result}
\end{figure*}


We illustrate and compare the generation results of our method and the baseline in Fig.~\ref{fig:knitwear_result}, where each group corresponds to a distinct requirement. For instance, the group (a4, b4) is generated based on the requirement \textit{“A desert-inspired knitted dress that evokes a sense of mystery and elegance”}. In Fig.~\ref{fig:knitwear_result}-(a4), our approach selects \texttt{bias-cut knit dress} in \textbf{garment type} to convey a sense of mystery, and \texttt{geomorphic} style in \textbf{aesthetic style} to reflect desert forms. For texture and surface expression, it chooses \texttt{striped knitted ribs} in \textbf{surface pattern} and \texttt{seed stitch} in \textbf{knitting technique}. To reinforce the theme, it uses \texttt{desert tones color} in \textbf{color palette} and \texttt{grain of shifting sand} in \textbf{visual motif} as visual inspiration. In contrast, the baseline result (Fig~\ref{fig:knitwear_result}-(b4)) renders a generic yellow dress set against a desert background, lacking the structural and stylistic richness of (a4) to express the intended sense of mystery. This highlights the advantage of design-space-guided generation in producing garments that align more closely with the intended concept. These examples highlight the advantage of design-space-guided generation in producing knitwear renderings that better reflect the intended concept. Our approach effectively selects design elements aligned with the input requirement, leveraging domain knowledge to ensure both visual distinctiveness and consistent quality across diverse requirements.

\subsubsection{User Study}
To evaluate the effectiveness of our approach, we conducted a user study comparing the knitwear designs generated by our method and the baseline. We prepared 10 design requirements as inputs and used both approaches to generate a total of 20 knitwear designs, grouped into 10 pairs. Each pair contained one design from each method, and the order within each pair was randomized.



\textbf{Procedure.} We recruited 50 participants (7 male) aged between 19 and 35 ($M = 23.54$, $SD = 2.48$), all with academic backgrounds in design. Each participant was shown 10 pairs of knitwear designs, one pair at a time. Participants were instructed to rate the quality of each design using a 5-point Likert scale and the measurements include \textit{texture and tactile quality}, \textit{alignment with design requirements}, and \textit{aesthetics and creativity}.

\begin{figure}
  \centering
  \includegraphics[width=\linewidth]{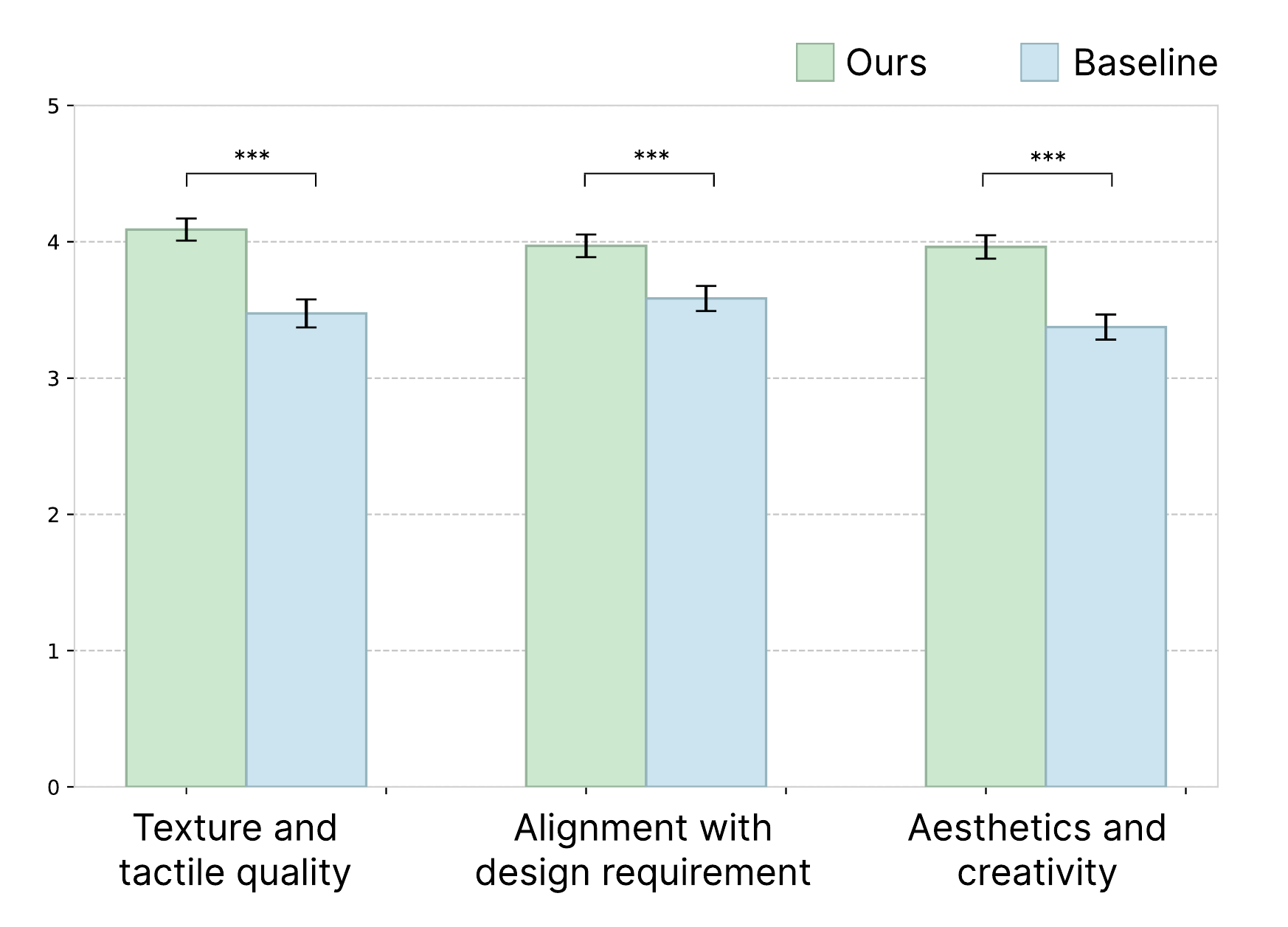}
  \caption{Means and 95\% confidence intervals for each measurement under our method and the baseline($\ast$: $p<.05$, $\ast\ast$: $p<.01$, $\ast\ast\ast$: $p<.001$).}
  \label{fig:knitwear_evaluation_results}
\end{figure}

\textbf{Results. } We conducted paired t-tests to compare our method with the baseline across all measurements. As shown in Fig.~\ref{fig:knitwear_evaluation_results}, our method received significantly higher ratings in all aspects. Specifically, for \textit{texture and tactile quality}, our designs ($M = 4.09$, $SD = 0.92$) were rated higher than the baseline ($M = 3.47$, $SD = 1.17$, $p < .001$). In terms of \textit{alignment with design requirements}, our method ($M = 3.97$, $SD = 0.94$) again outperformed the baseline ($M = 3.58$, $SD = 1.04$, $p < .001$). For \textit{aesthetics and creativity}, our approach ($M = 3.96$, $SD = 0.97$) also scored higher than the baseline ($M = 3.37$, $SD = 1.05$, $p < .001$). These results demonstrate that our approach generates knitwear designs with higher visual quality, conceptual alignment, and creative appeal.

\section{Limitations and Future Work}
In this section, we conclude several limitations and suggest potential solutions for future research.


\textbf{Automating Design Space Construction.} Manual construction of the design space is labor-intensive and demands domain expertise, which limits scalability across diverse design scenarios. Future work could investigate automatic methods by mining design repositories, extracting common patterns from existing artifacts, or leveraging large multimodal models. These strategies could reduce manual effort and improve adaptability to new design domains.

\textbf{Supporting Interactive Generation.} Current generation workflows take user requirements as input and directly output results. Users cannot view or adjust the generated constraints guiding the generation process, which limits both transparency and controllability. Subsequent studies may focus on visualizing the design space and its associated constraints, enabling users to inspect and refine them, thereby improving controllability and supporting more customized design generation.

\textbf{Handling Abstract Design Elements.} Our approach relies on a design space composed of concrete, composable elements like visual encodings that can be directly assembled through explicit operations. However, it struggles with abstract concepts such as moods or stylistic themes that lack explicit visual form. Future work could explore incorporating abstract design elements by mapping them to higher-level design strategies, thereby enabling implicit guidance during generation.

\textbf{Supporting Continuous Dimensions.} The current framework models the design space using categorical dimensions, where each element is selected from a discrete set. This limits its ability to represent continuous design parameters such as font size, padding, or color gradients. Future work could extend the formulation to incorporate continuous variables, enabling finer-grained control and expanding applicability to a broader range of design tasks.

\section{Conclusion}
This paper introduces CODS, a theoretical model that formalizes computational design as a constrained optimization process within a structured design space. By leveraging large language models to automatically generate design constraints from user requirements, CODS enables scalable, interpretable, and domain-independent design automation. Through applications in visualization and knitwear design, we demonstrate that CODS can produce high-quality, semantically aligned outcomes that outperform existing LLM-based approaches. Our work provides a unified and extensible foundation for intelligent design systems, paving the way for future research in interactive design, abstract design reasoning, and support for continuous design parameters.

\bibliographystyle{IEEEtran}
\bibliography{main}

\begin{IEEEbiography}[{\includegraphics[width=1in,height=1.25in,clip,keepaspectratio]{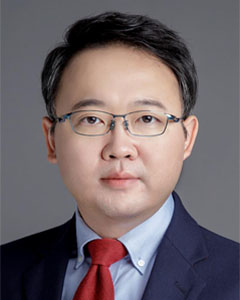}}]{Nan Cao}
received his Ph.D. degree in Computer Science and Engineering from the Hong Kong University of Science and Technology (HKUST), Hong Kong, China in 2012. He is currently a professor at Tongji University and the Vice Dean of the Tongji College of Design and Innovation. He also directs the Tongji Intelligent Big Data Visualization Lab (iDV$^x$ Lab) and conducts interdisciplinary research across multiple fields, including data visualization, human-computer interaction, and artificial intelligence. He was a research staff member at the IBM T.J. Watson Research Center, New York, NY, USA before joining the Tongji faculty in 2016.
\end{IEEEbiography}

\begin{IEEEbiography}[{\includegraphics[width=1in,height=1.25in,clip,keepaspectratio]{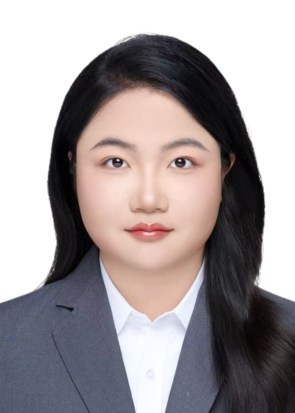}}]{Xiaoyu Qi}
received her bachelor's degree from the Department of Software Engineering, Tongji University in 2022. Currently, she is a master's candidate at Tongji University. Her research interests include AI-supported design and data visualization.
\end{IEEEbiography}

\begin{IEEEbiography}[{\includegraphics[width=1in,height=1.25in,clip,keepaspectratio]{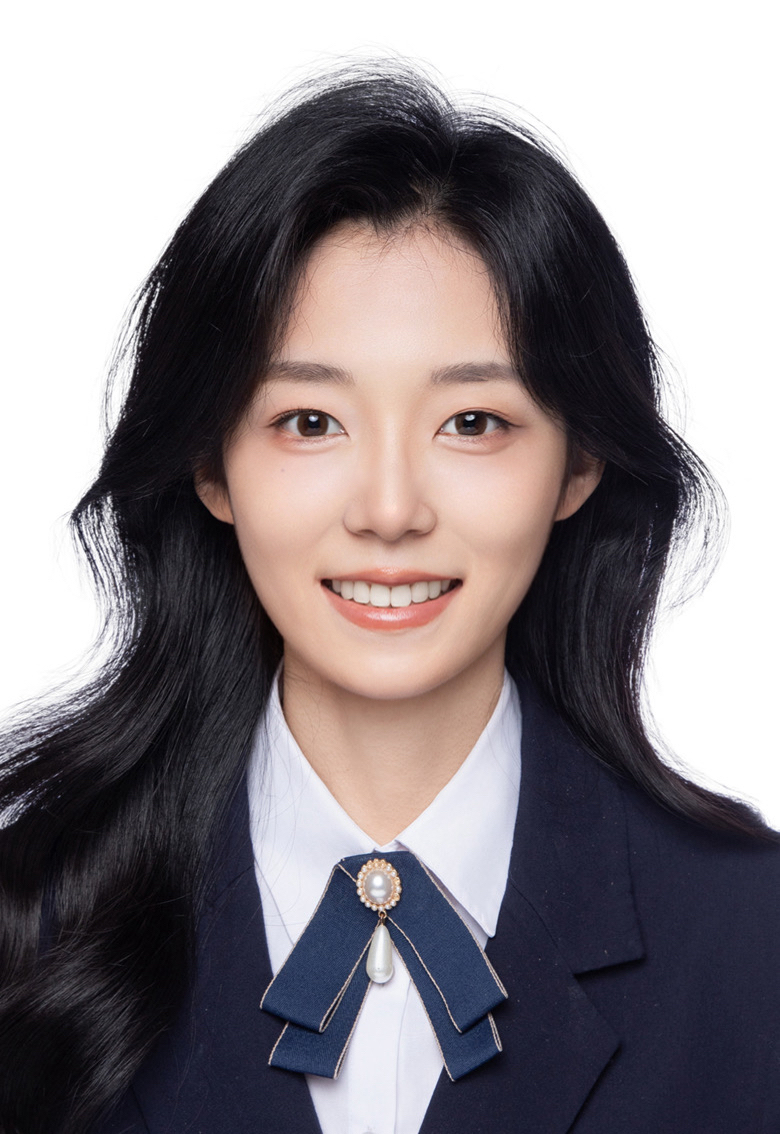}}]{Chuer Chen}
received her MSc degree from the Department of Electrical and Computer Engineering, National University of Singapore in 2021. She is currently working toward her Ph.D. degree as part of the Intelligent Big Data Visualization (iDVx) Lab, Tongji University. Her research interests include information visualization and intelligent design.
\end{IEEEbiography}

\begin{IEEEbiography}[{\includegraphics[width=1in,height=1.25in,clip,keepaspectratio]{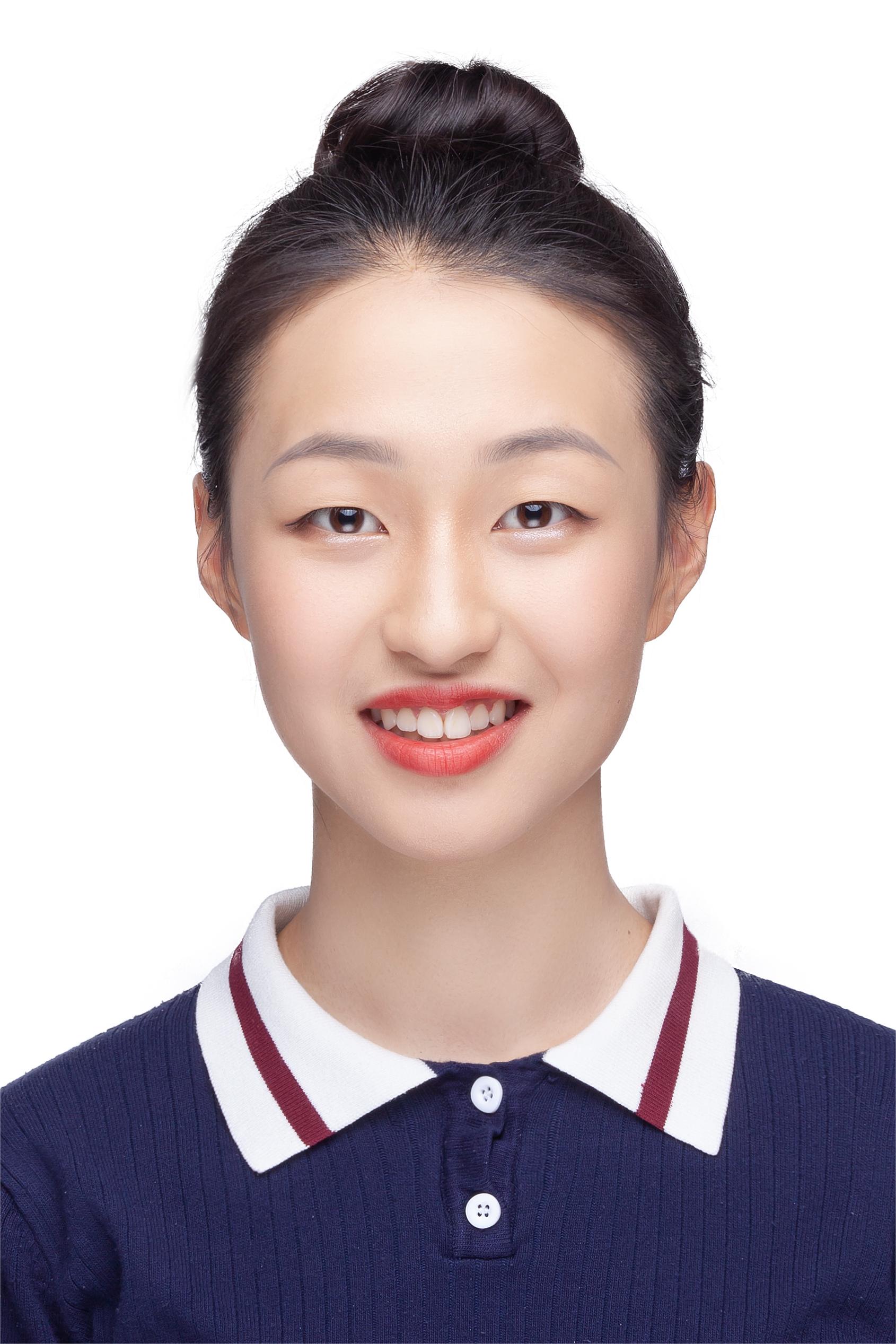}}]{Xiaoke Yan}
 received her bachelor’s degree from the Department of Software Engineering, Tongji University in 2024. She is currently pursuing her master’s degree at the College of Design and Innovation, Tongji University, where she is a member of the Intelligent Big Data Visualization Lab (iDVx Lab). Her research interests include information visualization, AI-supported design, and human-computer interaction.
\end{IEEEbiography}

\end{document}